\documentclass[prb,aps,twocolumn,superscriptaddress,showpacs]{revtex4}
\usepackage{epsfig}
\usepackage{amsmath}
\usepackage{amsthm}
\usepackage{amssymb,braket}
\usepackage{verbatim}
\usepackage{bm}
\usepackage{hyperref}
\usepackage[all]{hypcap}

\newcommand{\bpm}{\begin{pmatrix}}
\newcommand{\epm}{\end{pmatrix}}
\newcommand{\ba}{\begin{eqnarray}}
\newcommand{\ea}{\end{eqnarray}}

\newcommand{\smk}{\text{\tiny{\emph{K}}}}

\begin{document}
\title{Spin and Orbital Angular Momentum Structures of Cu(111) and Au(111) Surface States}

\author{Beomyoung Kim}
\affiliation{Institute of Physics and Applied Physics, Yonsei University, Seoul 120-749, Korea}

\author{Choong H. Kim}
\affiliation{Department of Physics and Astronomy, Seoul National University, Seoul 151-747, Korea}

\author{Panjin Kim}
\affiliation{Institute of Physics and Applied Physics, Yonsei University, Seoul 120-749, Korea}

\author{Wonsig Jung}
\affiliation{Institute of Physics and Applied Physics, Yonsei University, Seoul 120-749, Korea}

\author{Yeongkwan Kim}
\affiliation{Institute of Physics and Applied Physics, Yonsei University, Seoul 120-749, Korea}

\author{Yoonyoung Koh}
\affiliation{Institute of Physics and Applied Physics, Yonsei University, Seoul 120-749, Korea}

\author{Masashi Arita}
\affiliation{Hiroshima Synchrotron Radiation Center, Hiroshima University, Higashi-Hiroshima, Hiroshima 739-0046, Japan}

\author{Kenya Shimada}
\affiliation{Hiroshima Synchrotron Radiation Center, Hiroshima University, Higashi-Hiroshima, Hiroshima 739-0046, Japan}

\author{Hirofumi Namatame}
\affiliation{Hiroshima Synchrotron Radiation Center, Hiroshima University, Higashi-Hiroshima, Hiroshima 739-0046, Japan}

\author{Masaki Taniguchi}
\affiliation{Hiroshima Synchrotron Radiation Center, Hiroshima University, Higashi-Hiroshima, Hiroshima 739-0046, Japan}

\author{Jaejun Yu}
\affiliation{Department of Physics and Astronomy, Seoul National University, Seoul 151-747, Korea}

\author{Changyoung Kim}
\email[Electronic address:$~~$]{changyoung@yonsei.ac.kr}
\affiliation{Institute of Physics and Applied Physics, Yonsei University, Seoul 120-749, Korea}

\begin{abstract}
We performed angle resolved photoemission (ARPES) studies on Cu(111) and Au(111) surface states with circularly polarized light. Existence of local orbital angular momentum (OAM) is confirmed as has been predicted to be broadly present in a system with an inversion symmetry breaking (ISB).  The single band of Cu(111) surface states is found to have chiral OAM in spite of very small spin-orbit coupling (SOC) in Cu, which is consistent with theoretical prediction. As for Au(111), we observe split bands for which OAM for the inner and outer bands are parallel, unlike the Bi$_2$Se$_3$ case. We also performed first principles calculation and the results are found to be consistent with the experimental results. Moreover, majority of OAM is found to be from $d$-orbitals and a small contribution has $p$-orbital origin which is anti-aligned to the spins. We derive an effective Hamiltonian that incorporates the role of OAM and used it to extract the OAM and spin structures of surface states with various SOC strength. We discuss the evolution of angular momentum structures from pure OAM case to a strongly spin-orbit entangled state. We predict that the transition occurs through reversal of OAM direction at a $k$-point in the inner band if the system has a proper SOC strength.
\end{abstract}
\maketitle


\section{Introduction}

When solids possess both inversion and time reversal symmetries, Kramer's theorem dictates that each electronic state is doubly spin degenerate.\cite{kramers} When the inversion symmetry is broken at surfaces and interfaces, one can expect that the spin degeneracy is lifted except at some special $k$-space points. Lifting of spin degeneracy and resulting band splitting in two dimensional electron gas such as metallic surface states are typically explained in terms of the Rahsba effect.\cite{Rashba} In the Rashba effect, electron spin interacts with an effective magnetic field stemming from electron motion in a surface electric field and results in Zeeman splitting. Consequently, the surface state band splits and attains a chiral spin structure at a constant energy contour. Surface band splitting and concomitant chiral spin structure have been experimentally observed on surfaces of relative high atomic number metals\cite{Lashell,Osterwalder,Bi,Spin}, interfaces of hetero-structures\cite{Hetero} and surface states of topological insulators.\cite{Hasan}

In spite of its success in explaining the energy splitting and chiral spin structure, the original Rashba model could not provide the proper energy scale, giving about $10^5$ times smaller value than the measured one in the case of Au(111) surface bands. There have been several theoretical studies to resolve the issue but they did not address all the aspects of the Rashba effect.\cite{Petersen,Miki,Yaji,Bihlmayer,Frantzeskakis,Oguchi} It was only recently found that local orbital angular momentum (OAM) plays the key role in the Rashba-type band splitting by inducing asymmetric charge distribution in Bloch states when the atomic spin-orbit coupling (SOC) is much larger than the crystal field energy.\cite{SRParkPRL} The asymmetric charge distribution interacts with the surface electric field and provides the energy scale. It results in chiral OAM structure and chiral spin structure naturally follows from the strong SOC. A surprise came when chiral OAM was found to exist even if there is no SOC.\cite{JHPark} In this case, contrary to the strong SOC case, OAM vectors for the degenerate state are found to be parallel to each other while spins are anti-parallel.

As discussed above, the spin and OAM configurations and band splitting are quite different for the weak and strong SOC cases. Then, it would be interesting to investigate the transition from weak SOC to strong SOC cases. It is especially meaningful because the most studied Rashba split surface states on Au(111) possibly belong to an intermediate SOC case. To address this issue, we performed angle resolved photoemission (ARPES) experiments on Cu(111) and Au(111) surface states with circularly polarized light as well as first-principles calculation. We also develop an effective Hamiltonian to study the problem in an analytical way. We confirm that OAM is indeed present in the Cu(111) and Au(111) surface bands with most contribution from $d$-orbitals. Analysis based on the effective Hamiltonian shows that transition from parallel OAM in the weak SOC case to anti-parallel OAM in the strong SOC case occurs through reversal of OAM at a $k$-point in the inner surface band.

\section{Methods}

ARPES measurements were performed at the beam line 9A of HiSOR equipped with VG-SCIENTA R4000 analyzer. Data were taken with right and left circularly polarized (RCP and LCP) $10$ eV photons. The total energy resolution was set to be $10$ meV at $10$ eV, and the angular resolution was $0.1^{\circ}$. We performed experiment at $10$ K under a base pressure better than $7.5 \times  10^{-11}$ Torr. To obtain clean and well-ordered surfaces, we cleaned the surfaces by using Ar sputtering and annealed the samples by e-beam heating. In order to check high quality of the sample surface, we performed low energy electron diffraction (LEED) and confirmed a long range order. For the density-functional theory (DFT) calculations within the local-density approximation (LDA), we used the OpenMX\cite{openmx} based on the linear-combination-of-pseudo-atomic-orbitals (LCPAO) method.\cite{Ozaki} Spin-orbit interaction was included via the norm-conserving, fully relativistic $j$-dependent pseudopotential scheme in the non-collinear DFT formalism.\cite{Hill} To calculate the OAM and spin angular momentum (SAM) for a specific $k$-point, we used the LCAO coefficients of local atoms. Due to the non-orthogonality of pseudo-atomic-orbitals, LCAO coefficients are not strictly normalized to unity. To compensate for this, we re-normalized the coefficients by assuming the orthogonality, and obtained SAM and OAM values that are strictly bounded above by $0.5$ and $1$, respectively.

\section{Results and Discussion}
\begin{figure}
\centering \epsfxsize=8cm \epsfbox{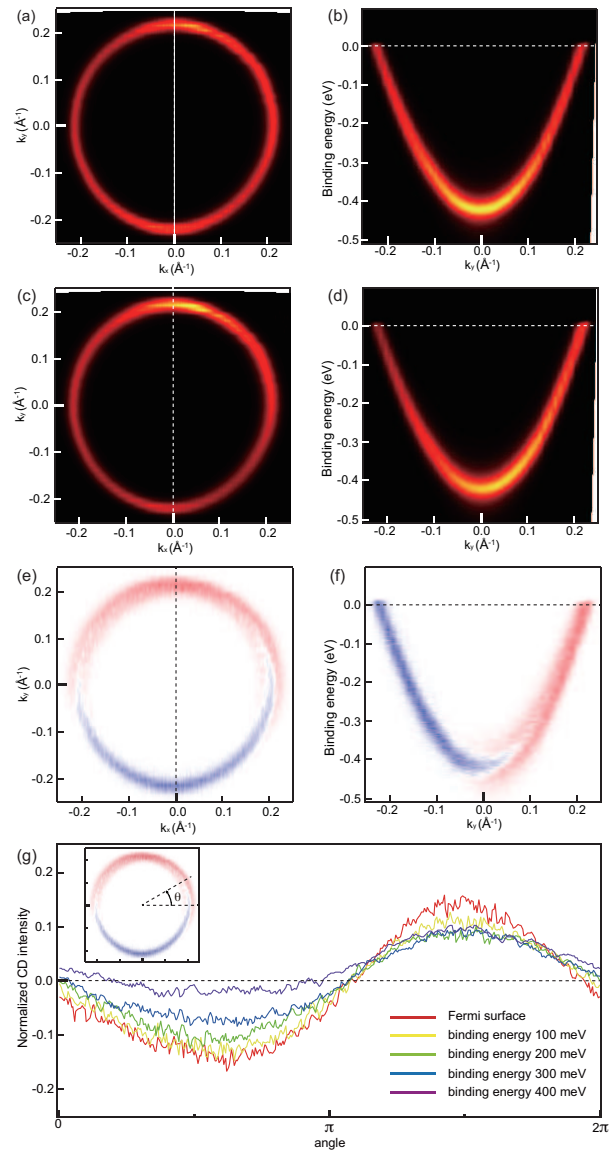} \caption{ARPES results from Cu(111) surface states. (a) Fermi surface and (b) the cut along the $k_x=0$ line (dashed line in panel (a)) taken with RCP light. (c) and (d) The same for LCP light. (e) and (f) RCP $-$ LCP data. (g) Normalized CD defined as NCD $=$ (RCP $-$ LCP) $/$ (RCP $+$ LCP) as a function of the azimuthal angle defined in the inset.}\label{fig1}
\end{figure}

In figure \ref{fig1}, we plot ARPES results from Cu(111) surface states. As expected, Cu(111) surface states  have spin degeneracy and show a single parabolic dispersive band due to the small atomic SOC. The binding energy at the   point is E($\Gamma) = 418$ meV, and the Fermi momentum $k_F = 0.211 \AA^{-1}$ (see figure. 1(a) - 1(d)). These values are consistent with published values.\cite{Mulazzi} In figure 1(e), we plot the circular dichroism (CD) at the Fermi energy defined as CD $=$ RCP $-$ LCP. The CD data presented in the color scale changes gradually from red to blue as value increases continuously from $-$ to $+$. Plotted in figure 1(f) is the cut image along $k_y=0$ shown as the dashed line in figure 1(e). This figure shows that CD is negative (positive) for positive (negative) $k_y$ for all energies. The band split near the point is an artifact resulted from broadening in the LCP data due to aging. It is clear from the raw data in figures 1(b) and 1(d) that there is only a single band. Finally, plotted in figure 1(g) is the normalized CD (defined as NCD $=$ (RCP $-$ LCP) $/$ (RCP $+$ LCP)) at constant binding energies as a function of the azimuthal angle defined in the inset. The curves have a sine function form, which suggests that the OAM forms a chiral structure.\cite{JHPark,SRPark} In addition, the estimated magnitude of OAM from CD decreases as the binding energy increases. The magnitude is found to be approximately proportional to the magnitude of the electron momentum value.

The behavior of CD in ARPES reveals that chiral OAM indeed exists in the Cu(111) surface band in spite that SOC is very small in Cu as predicted.\cite{JHPark} It is also consistent with the prediction that the magnitude of OAM is linear in the electron momentum value $k$. The formation of chiral OAM is a way of lowering the system energy by making the charge distribution asymmetric in the presence of surface electric field.\cite{SRParkPRL} As will be discussed later, OAM vectors for the degenerate band are parallel to each other while spins are anti-parallel. Note that if CD were due to spins, we would have not observed CD because there is no spin polarization for a state. The fact that we can observe CD from Cu(111) shows that CD is from OAM.\cite{SRPark,JHPark}

\begin{figure}
\centering \epsfxsize=8cm \epsfbox{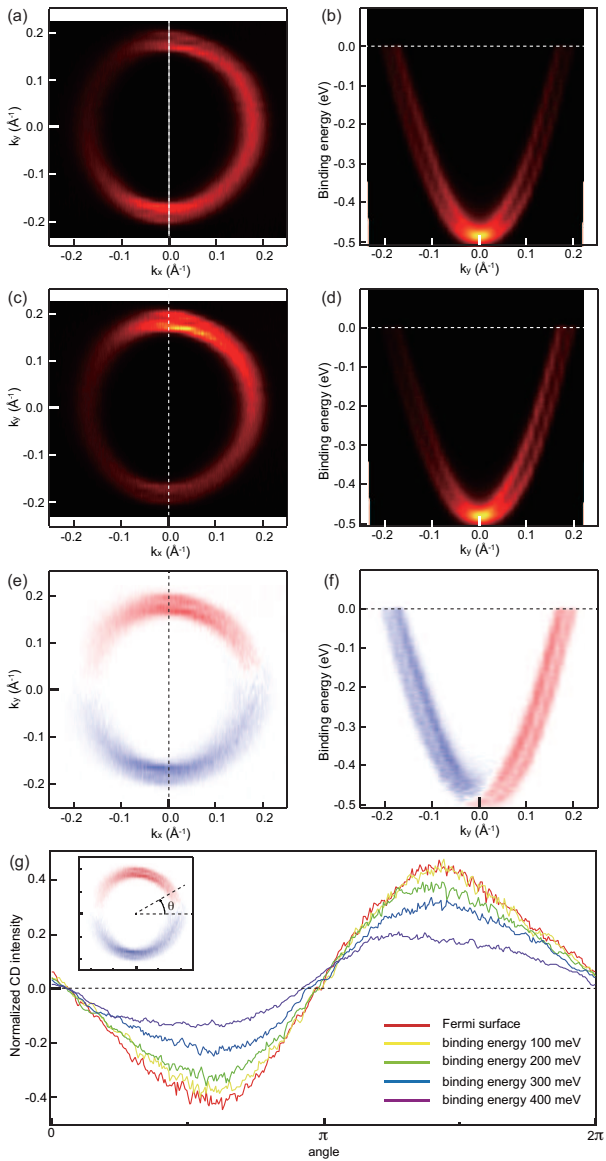}
\caption{ARPES results from Au(111) surface states. (a) Fermi surface and (b) the cut along the $k_x=0$ line (dashed line in panel (a)) taken with RCP light.  (c) and (d) The same for LCP light. (e) and (f) RCP $-$ LCP data. (g) NCD as a function of the azimuthal angle. Data from the two bands are summed in the estimation to have averaged NCD.}\label{fig2}
\end{figure}

We now turn our attention to the Au(111) surface states, possibly the most studied surface states in regards to the Rashba effect. The first direct experimental evidence for Rashba split bands was from the Au(111) surface.\cite{Lashell} This system has a similar band shape to that of Cu(111) surface states, but has band splitting of about 110 meV. In figure 2(a), we plot Fermi surface map of Au(111) surface states taken with RCP light. Clear double Fermi surfaces due to Rashba-type splitting are seen in the figure. Figure 2(b) shows the cut along the $k_x=0$ (dashed line in figure 2(a)). Once again, clear split bands are seen.  The observed band bottom is at E($\Gamma$) = 479 meV and the Fermi vectors are $k_F = 0.165 \AA^{-1}$ for the inner band and $k_F = 0.196 \AA^{-1}$ for the outer band. These values are quite consistent with reported values. Data taken with LCP light in figures 2(c) and 2(d) show similar features compared to the RCP data except there is some difference in the intensity profile.

Figures 2(e) and 2(f) plot circular dichroism RCP $-$ LCP for the Fermi surfaces and bands along the $k_x = 0$ line (dotted line in 2(e)). Except that there is clear band splitting, the overall CD profile looks similar to that of Cu(111) : it is negative (positive) for $k_y\textgreater 0 (\textless 0)$ region. Averaged NCD of the inner and outer bands as a function of the azimuthal angle $\theta$  plotted in figure 2(g) also shows a similar behavior to that of Cu(111) that it can be fit with a sine function and the magnitude is approximately linear in $k$. However, we also note that NCD value for Au is about three times larger than that of Cu, which suggests that OAM in Au case is generally larger than that for Cu. A peculiar aspect to note is that, unlike the Bi$_2$Se$_3$ case,\cite{SRPark} inner and outer bands have the same CD sign even though their spin directions are opposite. Same sign of CD suggests that OAM of the inner and outer bands are pointing in the same direction as in the Cu(111) case in spite of a stronger SOC and thus larger splitting. On the other hand, a careful look reveals that CD for the inner band is slightly stronger by about $35\%$ compared to that from the outer band.

In order to investigate the spin and OAM structures in more detail, we performed first principles density functional theory (DFT) calculations for Au(111) surface states within the local density approximation. In the DFT results plotted in figure 3, top panels (3(a) and 3(b)) show directions and sizes of spin and OAM by the arrows. The left panel 3(a) is for the inner band while the right one 3(b) is for the outer band. The spin and OAM directions obtained from DFT results are as expected. The directions for OAM (blue) and spin (red) are opposite to each other in the inner band while the outer band has parallel OAM and spin. The spin and OAM directions are consistent with previous report\cite{Lashell} and experimental result in figure 2, respectively. That is, while spins are anti-parallel between the two bands, OAM are pointing in the same direction.

We turn our attention to the magnitudes of OAM and spin. We first look at the spin magnitude. It is seen from the figures that the spin magnitudes from the two bands are quite similar (only the directions are opposite). Moreover, it has very little momentum dependence. In fact, the magnitude actually slightly decreases as the momentum increases. As for OAM, we find that OAM magnitude increases as we move away from the $\Gamma$ point, making OAM magnitude approximately linear in $k$ as indicated in the experimental results. However, the OAM magnitudes for the inner and outer bands are also very similar, which appears to contradict the experimental result.

\begin{figure}[ht]
\centering \epsfxsize=8cm \epsfbox{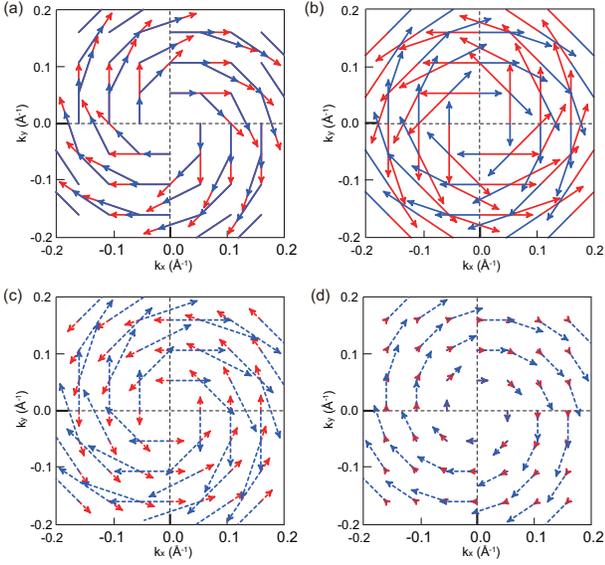} \caption{DFT results on the angular momentum structures of Au(111) surface states. DFT results for (a) inner and (b) outer bands. OAM (spin) of a state is represented by blue (red) arrow. Shown in the lower panels are contributions from $p$-state (red arrow) and $d$-state (blue dashed arrow) for (c) inner and (d) outer bands.}\label{fig3}
\end{figure}

To better understand the seemingly contradicting results from experiment and theory, we look into atomic orbital dependent contributions to OAM. For Au(111) surface states, 5$d$- and 6$p$-orbitals contribute to OAM\cite{Henk}. In figures 3(c) and 3(d), we plot contributions from 5$d$- and 6$p$-orbitals (blue and red, respectively). One can see that the $d$-orbital contribution dominates and determines the OAM direction. On the other hand, $p$-orbital contribution shows quite different behavior from that of $d$. The direction is opposite for inner and outer bands, making it always anti-parallel to the spin direction. Even though the total OAM is similar for the inner and outer bands, they are built differently and result in different circular dichroism. As a side note, the composition of the OAM shown in figures 3(c) and 3(d) suggests that the atomic SOC parameter for 6$p$ is larger than that of 5$d$ because a large value of atomic SOC parameter tends to anti-align the spin and OAM.

A natural question is how the spin and OAM structures evolve as a function of SOC strength, from a small SOC (e.g., Cu) to strong SOC (e.g. Bi$_2$Se$_3$). To elucidate the issue, we wish to develop an effective Hamiltonian and analyze the evolution. It has already been shown that free electron based model cannot explain various aspects of split bands and that tight binding state is needed.\cite{Petersen} A through derivation with nearest neighbor hoppings considered on tight binding states can be found elsewhere\cite{JHPark}. Instead, we wish to develop a simpler effective model for the surface states. For simplicity, we limit our discussion to the $p$-orbital case, but it can be extended to other orbitals.

For electrons in the surface states, there are 4 terms that are significant in the Hamiltonian. They are the kinetic energy $\hat{H}_K$, atomic spin orbit coupling $\hat{H}_{SOC}$, crystal field $\hat{H}_{CF}$, and the electrostatic energy due to interaction of asymmetric charge distribution with surface electric field $\hat{H}_{ES}$.\cite{SRParkPRL} For the kinetic energy $\hat{H}_K$, we simply add a $k^2$ term at the end to account for the free electron-like parabolic band. $\hat{H}_{SOC}$ is $\alpha \vec{L} \cdot \vec{S} $ where $\alpha$ is the atomic SOC parameter. Meanwhile $\hat{H}_{CF}$ is $\Delta$ for $p_x$ and $p_z$, and 0 for $p_y$ (note that we take $y$-axis as surface normal, not the usual $z$-axis). The last term $\hat{H}_{ES}$ comes from interaction of the surface electric field and electric dipole moment of a state with asymmetric charge distribution.\cite{SRParkPRL} The asymmetric charge distribution is a combined effect of electron momentum and local OAM, and should be proportional to the momentum and OAM. Therefore, the Hamiltonian is given by\cite{SRParkPRL,JHPark} $\hat{H}_{ES}=-\vec{p} \cdot \vec{E}_s = -\alpha_{\smk} (\vec{L} \times \vec{k})\cdot \vec{E_s}=-\alpha_{\smk} (\vec{k}\times\vec{E} _s)\cdot \vec{L}$ where $\alpha_{\smk}$ is a constant that is related to how efficiently asymmetric charge is created by $\vec{k}$ and $\vec{L}$. It is similar to the well known Rashba Hamiltonian $\hat{H}_R=\alpha_{\text{\tiny{R}}} (\vec{k}\times\vec{E} _s)\cdot\vec{\sigma}$ with the spin operator replaced by the OAM operator, but can account for the split energy.

We consider a state along the $x$-direction $\vec{k}=k_x\hat{x}$ without loss of generality, in which case $\hat{H}_{ES}=-\alpha_{\smk} k_x E _s \hat{L}_z$. The total Hamiltonian is estimated in the basis of $|$p$_{x\uparrow}\rangle$, $|$p$_{x\downarrow}\rangle$, $|$p$_{y\uparrow}\rangle$, $|$p$_{y\downarrow}\rangle$, $|$p$_{z\uparrow}\rangle$, and $|$p$_{z\downarrow}\rangle$. The result is given by

{\begin{align}\label{matrix}
    &\hat{H}= Ck^2\hat{I} + \nonumber\\
    &\left(
      \begin{array}{cccccc}
        \Delta & 0 & i\alpha_{\smk} k_x E_s-\frac{i\alpha}{2} & 0 & 0 & \frac{\alpha}{2} \\
         & \Delta & 0 & i\alpha_{\smk} k_x E_s+\frac{i\alpha}{2} & -\frac{\alpha}{2} & 0 \\
         &  & 0 & 0 & 0 & -\frac{i\alpha}{2} \\
         &  &  & 0 & -\frac{i\alpha}{2} & 0 \\
         &  &H.C.  &  & \Delta & 0 \\
         &  &  &  &  & \Delta \\
      \end{array} \nonumber
    \right)
\end{align}}
The 6$\times$6 matrix is diagonalized to obtain the eigen-states and energies. The two lowest energy states correspond to the surface states we have discussed above, and their spin and OAM can be easily estimated.

The spin and OAM of the two states are obtained as a function of atomic SOC parameter $\alpha$ to investigate the evolution of spin and OAM. The results are plotted in figure 4. Red (blue) color denotes spin (OAM) while dotted (solid) line is for the inner (outer) band. $\alpha$ increases as we go from left to right. For the small $\alpha$ case in figure 4(a), the band is degenerate. The OAM and spin structures are consistent with the results discussed above. OAM forms a chiral structure with the magnitude approximately linear in $k$ while spins are anti-parallel to each other but co-linear with OAM. This case represents the Cu(111) surface states. We also look at the other extreme of the large $\alpha$ case in figure 4(c). In this case, the degeneracy is lifted with a large band splitting. OAM and spin are always anti-parallel to each other due to the large $\alpha$ and the magnitude of OAM is independent of the electron momentum $k$, which is consistent with the results from Bi$_2$Se$_3$.\cite{SRPark}

\begin{figure}
\includegraphics[width=85mm]{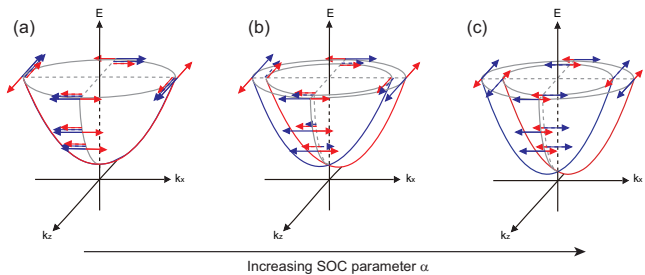}
\caption{Spin and OAM structures calculated by using the effective Hamiltonian for (a) small, (b) intermediate and (c) large atomic SOC parameter $\alpha$. All the parameters are fixed except $\alpha$. We chose 0.195 $\AA^{-1}$ as the Fermi momentum $k_F$ of the outer band, similar to the case of Au(111). The red and blue arrows represent spin and OAM, respectively. Dotted (solid) arrows are for the inner (outer) band.}\label{fig4}
\end{figure}

We finally examine the intermediate case in figure 4(b) which should provide us information on how parallel OAM for the small $\alpha$ case evolves to anti-parallel OAM for the large $\alpha$ case. For the outer band, OAM and spin are anti-parallel and the OAM magnitude has a linear $k$ dependence. This is similar to the small $\alpha$ case. The inner band, however, shows a quite different behavior. While OAM and spin are parallel to each other near the Fermi energy, they become anti-parallel near the $\Gamma$ point (small $k$ value). When $k$ increases from $0$, OAM gradually decreases, reverses the direction and increases again. Therefore, the OAM structure evolves from parallel configuration in the small $\alpha$ case to anti-parallel configuration in the large $\alpha$ case by reversing the OAM direction in the inner band at a certain $k$ point instead of reversing the direction at all $k$ points. Such behavior stems from the fact that $\hat{H}_{SOC}$ is larger than $\hat{H}_{ES}$ for a small $k$ and OAM prefers to stay anti-parallel to the spin. On the other hand, for a large $k$ value, $\hat{H}_{ES}$ is dominant and the system lowers the energy by inducing parallel OAM as in the Cu(111) case.  In the case of Au, such reversal of the OAM in the inner band was not observed, which means $\alpha$ for Au may not be large enough. Observation of such OAM reversal may be possible for surface states of Pb.


\section{Summary}

ARPES studies on Cu(111) and Au(111) surface states with circularly polarized light as well as first principles calculation have been performed to investigate the spin and OAM structures. Experimental and theoretical results show that OAM indeed exists even for Cu(111) and Au(111) surface states as predicted earlier.\cite{JHPark} Cu(111) has a degenerate single band with chiral and parallel OAM while Au(111) has split bands with OAM structure not too much different from that of Cu(111). DFT calculation shows that majority of OAM comes from $d$-states while a small contribution is from $p$-orbitals. We also developed an effective Hamiltonian with the role of OAM incorporated to investigate the evolution of the spin and OAM structures as a function of the atomic SOC parameter. We find that there should be OAM reversal at a specific momentum in the inner band when the system has a proper spin-orbit coupling strength.

This work is supported by NRF (contract No. 20090080739) and the KICOS under Grant No. K20602000008. This work was performed with the approval of the Proposal Assessing Committee of HSRC (Proposal No.10-A-57).

\end{document}